# A Theoretical Perspective on Transient Photovoltage and Charge Extraction Techniques

Oskar J. Sandberg,*,[†] Kristofer Tvingstedt,*,[§] Paul Meredith,[†] and Ardalan Armin*,[†]

[†] Sustainable Advanced Materials (Sêr-SAM), Department of Physics, Swansea University, Singleton Park, Swansea SA2 8PP Wales, United Kingdom

[§] Experimental Physics VI, Julius Maximillian University of Würzburg, 97074 Würzburg, Germany

Corresponding Authors

*E-mail: o.j.sandberg@swansea.ac.uk

*E-mail: ktvingstedt@physik.uni-wuerzburg.de

*E-mail: ardalan.armin@swansea.ac.uk

**Abstract**

Transient photovoltage (TPV) is a technique frequently used to determine charge carrier lifetimes in thin-film solar cells such as organic, dye sensitized and perovskite solar cells. As this lifetime is often incident light intensity dependent, its relevance to understanding the intrinsic properties of a photoactive material system as a material or device figure of merit has been questioned. To extract complete information on recombination dynamics, the TPV measurements are often performed in conjunction with charge extraction (CE) measurements, employed to determine the photo-generated charge carrier density and thereby the recombination rate constant and its order. In this communication, the underlying theory of TPV and CE is reviewed and expanded. Our theoretical findings are further solidified by numerical simulations and experiments on organic solar cells. We identify regimes of the open-circuit voltage within which accurate lifetimes and carrier densities can be determined with TPV and CE experiments. A wide range of steady-state light intensities is required in performing these experiments in order to identify their "working dynamic range" from which the recombination kinetics in thin-film solar cells can be determined.

# 1. Introduction

Thin-film devices based on organic and perovskite semiconductors hold the potential for the use within sustainable energy production[1-4] as well as other optoelectronic applications[5-9] such as sensors, photodetectors, and light-emitting diodes. In order to further improve and optimize the device performance over both short and long timescales, a better understanding of loss mechanisms such as recombination of the charge carriers is necessary.[10-12] For example, various different techniques, both electrical and optical, have been used to investigate the charge carrier recombination in thin-film solar cells. One of the more popular opto-electrical methods used to probe the recombination dynamics in this regard is the small-perturbation transient photovoltage technique (TPV), often measured in combination with charge extraction (CE) or differential charging/transient photocurrent.[13-16]

In TPV, the device under test is initially held at open-circuit under steady-state illumination.[13] To probe the recombination lifetime of the excess charge carriers, i.e. the carriers introduced into the device as a consequence of the steady-state light bias, an additional (weak) short light pulse is applied to the device and the corresponding induced voltage transient, developed over a large load resistance, is measured. The decay of this small photovoltage is then recorded from which the TPV lifetime is extracted. Ideally, this lifetime is associated with the steady-state recombination rate of charge carriers within the bulk. The small light perturbation increases the excess carrier density in the device as $n = n_{\text{oc}} + \delta n$, giving rise to a subsequent perturbation in the recombination rate $\mathcal{R} = \mathcal{R}_0 + \delta \mathcal{R}$; here, $n_{\text{oc}}$ is the photo-generated steady-state (dc) carrier density and $\mathcal{R}_0$ is the associated (time-independent) recombination rate, which is balanced by an equal steady-state photo-generation rate $G_0$ of charge carriers at open-circuit. The decay of the voltage transient, induced as a result of the small perturbation, is approximately given by[13]

$$\frac{d\Delta V(t)}{dt} \propto \frac{dn(t)}{dt} = -\delta \mathcal{R} = -\frac{\delta n}{\tau_B} \qquad (1)$$

leading to a first-order kinetics of $\Delta V(t) \propto \exp(-t/\tau_B)$, where $\tau_B$ is the associated pseudo-first-order bulk recombination lifetime.

In general, the open-circuit voltage is related to the steady-state light intensity or photo-generation rate via $V_{\text{oc}} \propto (n_{\text{id}} kT/q) \ln(G_0)$, where $n_{\text{id}}$ is the diode ideality factor and $kT/q$ is the thermal voltage.[17-19] Since the open circuit voltage is logarithmically dependent on the steady-state (dc) photo-generated carrier density, the recombination lifetime from TPV measurements is parametrically dependent on the $V_{oc}$, taking the form[13,19]

$$\tau_B = \tau_0 \exp\left(-\frac{qV_{oc}}{\nu kT}\right) \tag{2}$$

where $\nu$ is the slope parameter that ideally depends only on the order of the steady-state recombination dynamics within the bulk, while $\tau_0$ is a voltage-independent prefactor. For example, in the case of second-order bulk recombination $\mathcal{R} = \beta n^2$, the carrier recombination dynamics after the perturbation pulse is given by

$$\frac{dn}{dt} = G_0 - \beta(n_{oc} + \delta n)^2 \approx -\frac{\delta n}{\tau_B} \tag{3}$$

For $\delta n \ll n_{oc}$ the bulk recombination lifetime is thus

$$\tau_B = \frac{1}{2\beta n_{oc}} = \frac{1}{2\sqrt{\beta G_0}} \tag{4}$$

Subsequently, with $n_{id} = 1$, the open-circuit voltage dependence of Eq. (2) corresponds to $\nu = 2$ in this case. Here, $\beta$ is the second-order recombination coefficient and $n_{oc} = \sqrt{G_0/\beta}$ (since $G_0 = \beta n_{oc}^2$ at open-circuit conditions). This can be compared to the first order recombination dynamics scenario, occurring for example in heavily doped solar cells under lower illumination conditions where excess carrier density is smaller than the doping level. Then, $\nu = \infty$ and the lifetime is constant, i.e. independent of the concentration of photo-generated carriers.

To obtain the recombination rate and its kinetics from the TPV lifetime at different open-circuit voltages (i.e. different steady-state light intensities), the corresponding steady-state charge carrier density needs to be known as well. However, the exact relation between the steady-state carrier density $n_{oc}$ (at open-circuit) and the open-circuit voltage is unknown due to the system-specific voltage dependence of the recombination losses. As such, it is never possible to directly convert a $V_{oc}$ value to a charge carrier density, but the relationship needs instead to be measured as accurately as possible. To probe the carrier lifetime as a function of the photo-induced carrier density in the active layer, the transient photovoltage technique is therefore typically performed in combination with charge extraction (CE) or differential charging/transient photocurrent measurements. The ability of both of these methods to correctly determine bulk carrier concentrations is heavily debated, both in the organic and perovskite solar cell community, but are both frequently used to assign the carrier density as a function of open-circuit voltage. The extracted carrier density under open-circuit conditions is generally taken to follow an exponential open-circuit voltage dependence according to:

$$n_{\text{CE}} = n_{CE,0} \exp\left(\frac{qV_{oc}}{mkT}\right) \quad (5)$$

where $m$ is the important (and unknown) associated slope parameter, and $n_{CE,0}$ is a prefactor.[19] Measuring CE in conjunction with TPV allows for the recombination rate $R \sim n_{\text{CE}}/\tau_{TPV}$ to be mapped as a function of carrier density, providing the sought-after information concerning the dominating recombination mechanism.

In accordance with Eq. (1), one expects the lifetime extracted from the TPV experiment to reflect the recombination of photo-induced carriers. However, this interpretation has been questioned by Street who suggested that the extracted TPV lifetimes in organic solar cells is set by the device capacitance and given by[20,21]

$$\tau = C \left[\frac{\partial J_{\text{dark}}}{\partial V}\right]^{-1} \approx \frac{n_{\text{id}}kTC}{qJ_{\text{dark}}(V_{\text{oc}})} \quad (6)$$

where $J_{\text{dark}}$ is the dark steady-state current density and $C$ is the capacitance of the active layer. This "lifetime" can therefore be interpreted as an internal discharging time of the diode itself rather than reflecting the actual recombination dynamics of photo-induced carriers. Similar concerns on lifetime assignments obtained via voltage transient methods were raised for silicon solar cells in 1981 by Mahan and Barnes and analytically explained in more detail by Castener.[22,23] Recently, Kiermasch et al. thus proposed the TPV lifetimes to instead be given by the sum of the real recombination lifetime $\tau_B$ of photo-induced carriers and a capacitive contribution[24]

$$\tau_{\text{total}} = \tau_B + \frac{n_{\text{id}}kTC}{qJ_0} \exp\left(-\frac{qV_{\text{oc}}}{n_{\text{id}}kT}\right) \quad (7)$$

where $J_0$ is the dark saturation current density of the device. The capacitive contribution, corresponding to the second term on the right-hand side of Eq. (7) and dominating at lower voltages, is equivalent to the internal diode discharging lifetime suggested by Street.

In the same work, Kiermasch and co-workers also found the charge density (Eq. (5)), as extracted from CE, to be strongly influenced by capacitive effects, in particular at low voltages, leading to unrealistically large values for the slope parameter $m$ at these intensities.[24] As a consequence, using the charge carrier density obtained from CE in combination with the lifetimes extracted from TPV to probe the underlying recombination reaction order might often result in recombination orders > 2, which do not necessarily reflect the recombination of steady-state photo-induced carriers within the bulk. These type of higher recombination orders have usually been explained in terms of trap-assisted recombination via localized tail states, both in organic[25-27] and perovskite[28] solar cells. However, it has also been pointed out that higher recombination orders might be artefacts caused by non-uniform carrier profiles in thin devices.[19,29]

Given this background, and the recent insights of Kiermasch et al., in this work we further elaborate the limitations of TPV and CE and their physical meaning, from a theoretical view point supported by numerical calculations and demonstrative experiments. Starting from fundamental electrical transient current theory in the time-domain, we derive analytical expressions for the TPV lifetime and extracted charge carrier densities from CE. The validity of the analysis is confirmed with numerical drift-diffusion simulations on organic thin-film solar cell devices.

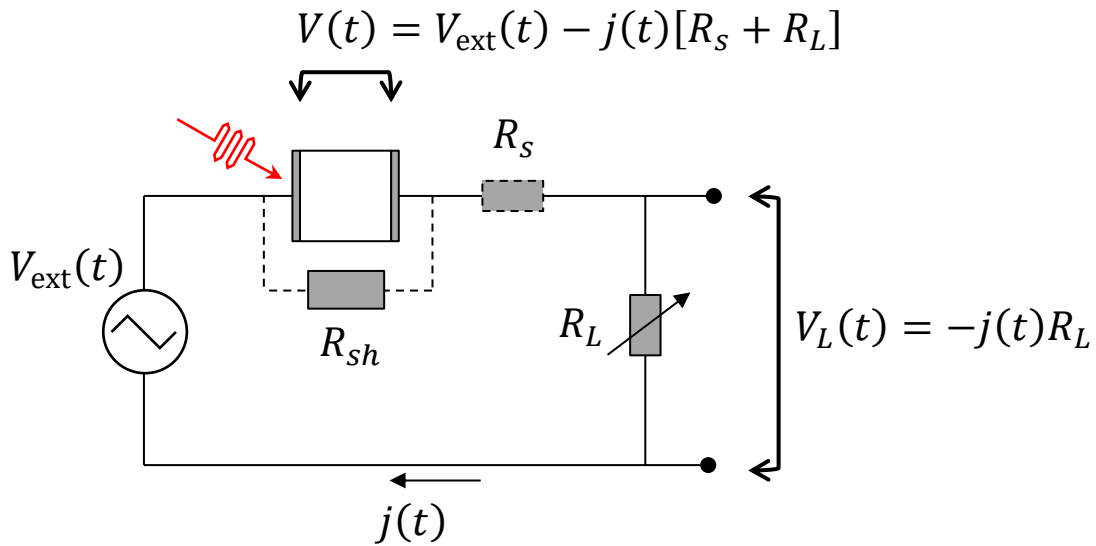

**Figure 1.** The general schematic set up for electrical transient measurements, such as TPV and CE, on diode or solar cell devices. The output voltage $V_L$ (or current $j$) is measured across the load resistance $R_L$ (usually via an oscilloscope). Note that in order to establish open-circuit conditions, the external voltage source is in general not needed, as long as a large enough load resistance is used; the case without the external voltage source corresponds to $V_{ext} = 0$.

## 2. Theory

A schematic picture of the electrical set up used for transient measurements such as TPV and CE is shown in Figure 1. In general, for a planar current flow across the (non-ideal) diode or solar cell device, the transient current density is given by[30,31]

$$j(t) = \bar{J}_c(t) + \frac{\varepsilon \varepsilon_0}{d} \frac{\partial V}{\partial t} \tag{8}$$

where $\bar{J}_c(t) \equiv (1/d) \int_0^d J_c(x,t)\, dx$ is the average conduction current density in the active layer and the second term on the right-hand side is the corresponding average displacement current density, with $V(t)$ being the voltage across the active layer:

$$V(t) = V_{\text{ext}}(t) - [R_L + R_s]j(t) \tag{9}$$

where $V_{\text{ext}}(t)$ is the external applied voltage (by a voltage source), $R_L$ is the load resistance (input terminal to the oscilloscope), and $R_s$ is the total series resistance of the electrodes, the external wires, and the internal resistance of the voltage source. In the set up shown in Figure 1, the transient voltage or current is obtained by measuring the voltage drop $V_L(t) = -j(t)R_L$ across the load. Note that, *under steady-state illumination*, open-circuit conditions can either be established by using a large $R_L$ ($R_L \to \infty$) and/or by applying an external voltage $V_{\text{ext}}$ equal to the corresponding open-circuit voltage.

In conjunction with the Poisson and the charge carrier continuity equations, the average conduction current can generally be expressed as[32]

$$\bar{J}_c(t) = qd\frac{\partial n_{\text{eff}}}{\partial t} + J_D(t) \tag{10}$$

as shown in the Supporting Information, with $n_{\text{eff}}$ being an effective charge carrier density in the device,

$$n_{\text{eff}} \equiv \frac{1}{2}[\bar{p} + \bar{n}] + \frac{\sigma'_{\text{el}}}{qd} \tag{11}$$

where $\bar{p}$ and $\bar{n}$ are the spatial averages of the total hole and electron carrier densities across the active layer, respectively, while

$$\sigma'_{\text{el}} = \frac{1}{d}\int_0^d \left(x - \frac{d}{2}\right)\rho(x,t)\, dx \tag{12}$$

where $\rho(x,t)$ is the total space charge density in the active layer. $\sigma'_{\text{el}}$ can be interpreted as an additional electrode charge induced by spatially *non-uniform* space charge distributions inside the active layer. Finally, $J_D$ is the sum of the net recombination-generation current density and the leakage current due to parasitic shunt resistance $R_{sh}$ (in $\Omega m^2$);

$$J_D(t) = q \int_0^d [\mathcal{R}_B(x,t) - G(x,t)] \, dx + J_{surf}(t) + \frac{V(t)}{R_{sh}} \tag{13}$$

where $G(x,t)$ is the photo-generation rate of free charge carriers, $\mathcal{R}_B(x,t)$ is the carrier recombination rate in the bulk, and $J_{surf}(t) = J_n(0,t) + J_p(d,t)$ is the total surface recombination current of electrons at the anode ($x = 0$) and holes at the cathode ($x = d$); $J_{n/p}(x,t)$ is the local electron/hole current density at the position $x$. We note that, in general, $J_D$ (or $J_{surf}$) also includes a contribution from Ohmic conduction currents (e.g. in case of a doped layer).

The above set of equations generally describes planar (one-dimensional) electrical current transients in sandwich-type devices. It should be noted that under dc conditions ($\partial/\partial t = 0$), the total current density reduces to $J_D$, taking its steady-state value. In the dark, this current density can usually be approximated as

$$J_D|_{dark} = J_0 \left[ \exp\left(\frac{qV}{n_{id}kT}\right) - 1 \right] + \frac{V}{R_{sh}} \tag{14}$$

which is the (non-ideal) generalized Shockley diode equation. Here, the applied voltage across the diode under dc conditions is given by $V = V_{ext} - J_D R_{s,tot}$, with $R_{s,tot} = R_L + R_s$ being the total series resistance of the external circuit (see Figure 1). Note, however, that in practice, the dark dc current is always measured without a separate load resistor so that $V = V_{ext} - J_D R_s$ in Eq. (14). In the following, the general theory [Eqs. (8) to (13)] is applied to TPV and CE relevant conditions.

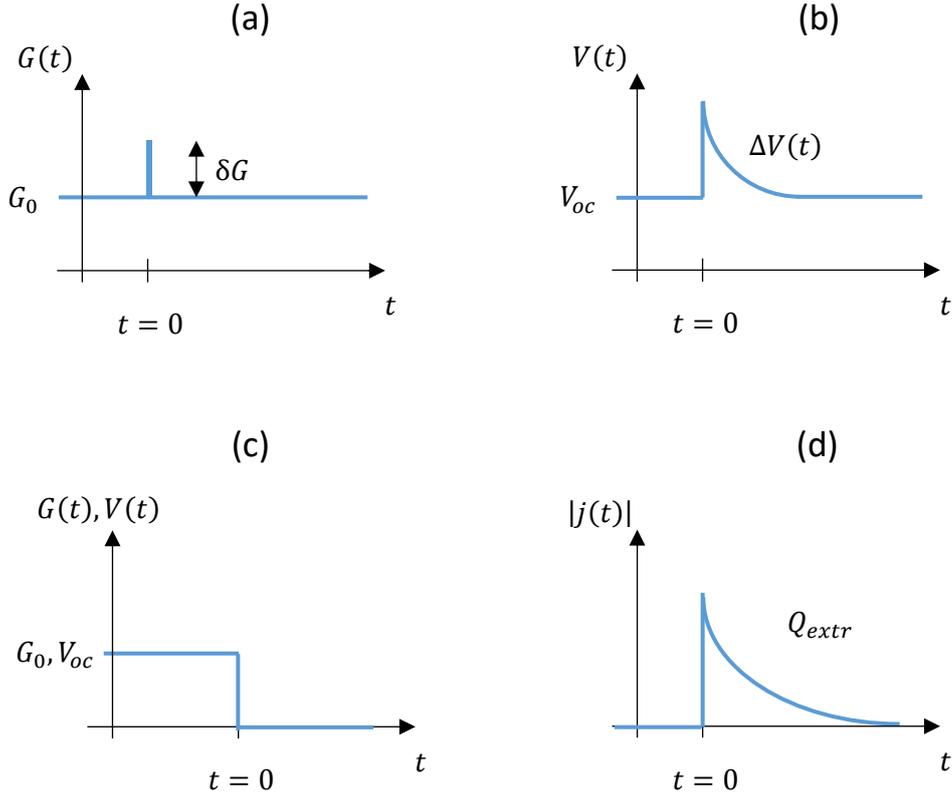

**Figure 2.** Schematic picture of TPV and CE. In TPV, a perturbation light pulse is applied under open-circuit voltage (a), giving rise to an induced voltage transient across the device (b). From the exponential decay of this transient, the TPV lifetime is extracted. In CE, the device initially held under illumination at dc open-circuit conditions is short-circuited with the light simultaneously switched off (c), this give rise to an extraction current transient flowing through the external circuit (d). The extracted charge is obtained by integrating the current transient response.

## 2.1. Small-Perturbation Transient Photovoltage

In TPV, a large load resistance ($R_\text{L} \gg R_\text{s}$) is connected in series with the device. This is to maintain open-circuit conditions during the perturbation light pulse; in practice, a load resistance of 1 MΩ is commonly used, although it is preferable to use much larger resistances to avoid shunting over the measurement load at low voltages.[33] From the (ideally) exponential decay of the voltage transient, the associated small perturbation TPV lifetime is determined. A schematic picture is illustrated in Figure 2a and b. Prior to the light perturbation, we have $V = V_\text{ext} - jR_\text{L} = V_\text{ext} + V_{L,0} = V_\text{oc}$, where $V_\text{oc}$ is the steady-state (time-independent) open-circuit voltage due to the background light and $V_{L,0}$ is the corresponding steady-state voltage drop across the load. In this case, the only function of applying a *fixed* external voltage $V_{ext}$ is to set the value for $V_{L,0}$, leaving the voltage $V_{oc}$ across the active layer unchanged. For example, if $V_{ext} = 0$ (i.e. no voltage source), then $V_{L,0} = V_{oc}$; conversely, if $V_{ext} = V_{oc}$, the steady-state voltage across the load is zero ($V_{L,0} = 0$).

Following the light perturbation, the photo-induced voltage transient over the active layer is given by $V(t) = V_{oc} + \Delta V(t)$, where $\Delta V(t) = V_L(t) - V_{L,0}$ is the small additional voltage drop induced over the load. Inserting Eq. (9) and (10) into Eq. (8), and rewriting the current equation in terms of $\Delta V(t)$, then reveals

$$C_0 \frac{\partial \Delta V}{\partial t} + \frac{\Delta V(t)}{R_L} = -qd \frac{\partial n_{eff}}{\partial t} - J_D \tag{15}$$

where $C_0 = \varepsilon \varepsilon_0/d$ is the geometric capacitance per unit area of the active layer. Provided that the perturbation is small, also the change in the current $J_D$ will be small and

$$J_D(V_{oc} + \Delta V) \approx \frac{\partial J_D(V_{oc})}{\partial V} \Delta V, \tag{16}$$

since $J_D(V_{oc}) = 0$ (open-circuit condition). Making use of the chain rule, Eq. (15) can then be re-expressed as:

$$\frac{\partial \Delta V}{\partial t} = -\left[\frac{\frac{1}{R_L} + \frac{\partial J_D(V_{oc})}{\partial V}}{C_0 + qd\frac{\partial n_{eff}(V_{oc})}{\partial V}}\right] \Delta V(t) \equiv -\frac{\Delta V(t)}{\tau_{TPV}} \tag{17}$$

where $\tau_{TPV}$ is the associated pseudo-first-order lifetime of the small-perturbation transient photovoltage decay.

The change in the effective charge density can be related to the recombination rate via $\delta n_{eff} = [\partial n_{eff}/\partial \bar{\mathcal{R}}]\delta\bar{\mathcal{R}}$, where $\bar{\mathcal{R}} = \bar{\mathcal{R}}_B + J_{surf}/qd$ is the total recombination rate, averaged over the active layer, incorporating both surface and bulk recombination; $\bar{\mathcal{R}}_B = (1/d)\int_0^d \mathcal{R}_B(x)\, dx$ is the average carrier recombination rate inside the bulk. On the other hand, making use of Eq. (13), we see that $qd\delta\bar{\mathcal{R}} = qd[\partial\bar{\mathcal{R}}/\partial V]\delta V = [\partial J_D/\partial V - 1/R_{sh}]\delta V$. Concomitantly, the TPV lifetime in Eq. (17) is equivalent to

$$\tau_{TPV} = (\tau_{B,eff} + R_0 C_0)\left[1 + \left(\frac{1}{R_{sh}} + \frac{1}{R_L}\right)R_0\right]^{-1} \tag{18}$$

where $\tau_{B,eff} = [\partial\bar{\mathcal{R}}/\partial n_{eff}]^{-1}$ is an effective carrier recombination lifetime, and

$$R_0 = \left[\frac{\partial J_D(V_{oc})}{\partial V} - \frac{1}{R_{sh}}\right]^{-1} \approx \frac{n_{id}kT}{qJ_0}\exp\left(-\frac{qV_{oc}}{n_{id}kT}\right) \qquad (19)$$

is the differential resistance of the diode itself (excluding the shunt resistance), where Eq. (14) was used in the last step. Note that the shunt resistance $R_{sh}$ plays an identical role to the load resistance $R_L$ in Eq. (18), consistent with previous experimental findings.[24,33]

In the limit when the electron and hole densities are homogenous across the active layer, corresponding to sufficiently high steady-state light intensities, we obtain $\tau_{B,\text{eff}} = \partial[(p+n)/2]/\partial\bar{\mathcal{R}}$, becoming identical to $\tau_B$ in Eq. (1) when $\bar{p} = \bar{n}$. Under these conditions and provided that $\tau_{B,\text{eff}} \gg R_0 C_0$ (and $R_0 \ll R_L, R_{sh}$), $\tau_{TPV}$ thus embodies a real carrier lifetime which is also relevant under steady-state conditions. Note that this lifetime is only a constant in the case of true first order recombination, but the general assumption is that $\tau_B$ is allowed to be dependent on the carrier concentration and should thus represent recombination dynamics of any order. We also note that for the case $n_{\text{eff}} = \bar{p} = \bar{n}$, Eq. (18) becomes equivalent to the TPV lifetime expression previously proposed by Credgington et al.[34,35], and can be simplified as $\tau_{TPV} = [1 + C_0/(\partial[q\bar{n}d]/\partial V)]\tau_B$ when external resistance effects are negligible.

At lower light intensities, however, it is expected that the carrier distributions inside the active layer (eventually) becomes very inhomogeneous. Under these conditions ($n_{\text{eff}} \neq \bar{n}, \bar{p}$), the effective lifetime $\tau_{B,\text{eff}} = [\partial\bar{\mathcal{R}}/\partial n_{\text{eff}}]^{-1}$ instead reflects how the total *effective* carrier density $n_{\text{eff}}$ changes with the steady-state recombination current. In the limit of small light intensities, the distribution of *excess* carriers is almost fully dominated by the *distribution of back-injected* dark carriers. In this case, the TPV lifetime in Eq. (17) can instead be approximated by

$$\tau_{\text{cap}} = \frac{C}{\left[\frac{\partial J_D(V_{oc})}{\partial V} + \frac{1}{R_L}\right]} = \left[\frac{1}{R_0} + \frac{1}{R_{sh}} + \frac{1}{R_L}\right]^{-1} C \qquad (20)$$

where

$$C = C_0 + qd\frac{\partial}{\partial V}[n_{\text{eff}}] \qquad (21)$$

is given by the *dark* capacitance of the active layer, *assuming* $n_{\text{eff}}$ to be governed by the dark back-injected carriers. In this limit, TPV is limited by a composite RC-time constant of the device, being ruled by the smallest of the load resistance, the shunt resistance, and the differential resistance of the diode itself.

We note that the RC decay in the limit of low light intensities is not fully determined by the geometric capacitance $C_0$, but rather by a total capacitance $C$, as given by Eq. (21). This total capacitance is given by the sum of $C_0$ and a space charge term that monitors the change in the background charge carrier distribution (as induced by the voltage perturbation) during the voltage transient. Only when the change in the space charge distribution is negligible, will $C$ be given by the geometric capacitance $C_0$. However, for the case of a doped active layer, for example, the photo-induced voltage change across the active layer will induce a small change in the (doping-induced) depletion layer thickness $w$ as well, and Eq. (20) reduces in this case to the depletion layer capacitance $C = \varepsilon\varepsilon_0/w$, as shown in the Supporting Information.

## 2.2. Charge Extraction

A schematic picture of the CE method is shown in Figure 2c and d. The device is initially held under illumination at open-circuit conditions. The device is then short-circuited (or reverse-biased) with the illumination switched off simultaneously, and the induced extraction current transient is measured. In contrast to TPV, the load resistance in CE needs to be as small as possible to avoid resistive losses; in the following, we assume the total series resistance $R_{\text{s,tot}} = R_\text{L} + R_\text{s}$ to be negligibly small (i.e. $jR_{\text{s,tot}} \ll V_\text{oc}$). The extracted charge, obtained by integrating the conduction current and correcting for the geometric capacitive electrode charge, is then given by

$$Q_{\text{CE}} \equiv -\left[\int_0^{t_\infty} j(t)\,dt - C_0 V_\text{oc}\right] = -\int_0^{t_\infty} \bar{J}_\text{c}(t)\,dt \qquad (22)$$

where $C_0 V_\text{oc}$ corresponds to the charge stored on the electrodes and $t_\infty$ is the time when all excess charge carriers have been collected (and/or recombined), so that $\bar{J}_\text{c}(t_\infty) = 0$. Inserting Eq. (10) into Eq. (22) and integrating, we obtain $Q_{\text{CE}} = q n_{\text{CE}} d$ with the associated extracted carrier density $n_{\text{CE}}$ given by

$$n_{\text{CE}} = \Delta n_{\text{eff}} - \int_0^{t_\infty} \bar{\mathcal{R}}\,dt \qquad (23)$$

where $\Delta n_{\text{eff}} = n_{\text{eff}}(t=0) - n_{\text{eff}}(t=t_\infty)$ and assuming shunt effects to be negligible.

When interpreting charge extraction experiments, the recombination between charge carriers during the charge extraction process of the excess carriers is generally assumed to be negligible

(complete charge extraction),[36] $\mathcal{R} = 0$. With this assumption, after making use of Eq. (11) and Eq. (12), Eq. (23) can be expressed as

$$n_{CE} = \frac{1}{d}\int_0^d \left[\frac{x}{d}\Delta p(x) + \left(1 - \frac{x}{d}\right)\Delta n(x)\right] dx \qquad (24)$$

assuming that $\Delta \rho(x) = q[\Delta p(x) - \Delta n(x)]$, where $\Delta p(x) = p(x, t = 0) - p(x, t = t_\infty)$ and $\Delta n(x) = n(x, t = 0) - n(x, t = t_\infty)$ are the corresponding excess hole and electron densities, respectively. Concomitantly, even in the ideal case of complete charge extraction, the extracted charge carrier density is, in general, not given by the spatial average of the extracted electron and hole densities within the active layer.

Therefore, in order to relate the extracted carrier density $n_{CE}$ to the steady-state carrier density under open-circuit conditions, the excess carrier distributions are also assumed to be homogenous and uniform throughout the active layer (i.e. $\Delta \sigma'_{el} = 0$). With this additional (implicit) assumption, Eq. (24) is then finally simplified as

$$n_{CE} = \frac{[\Delta p + \Delta n]}{2} = n_{oc} \qquad (25)$$

where $\Delta p = \Delta n = n_{oc}$ is the excess steady-state carrier density at open-circuit conditions. Note that in case of pure second-order recombination, we then expect $n_{CE} = n_{oc} = \sqrt{G_0/\beta}$, as before.

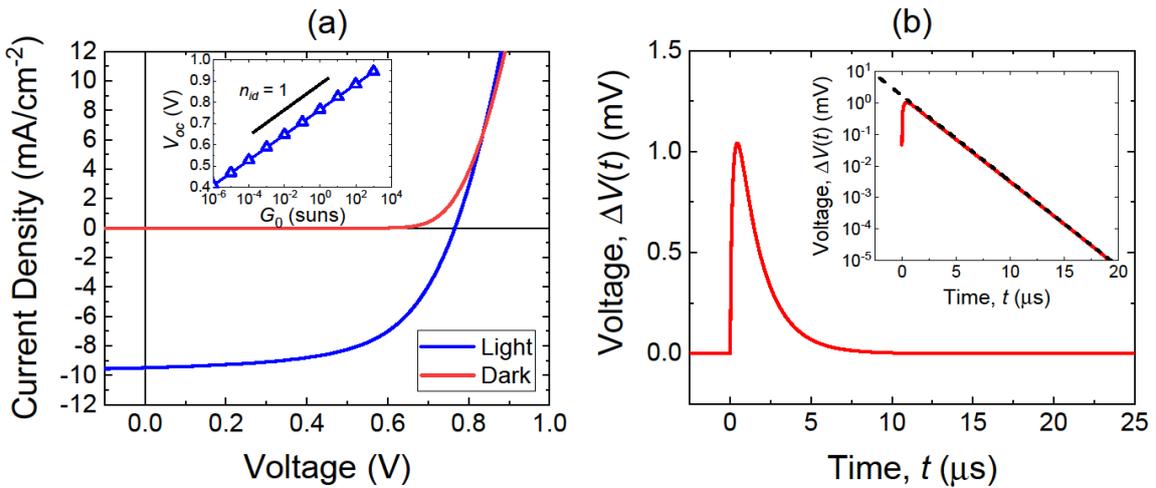

**Figure 3.** (a) The simulated current-voltage characteristics of an organic solar cell under consideration at 1 sun. In the inset, the open-circuit voltage is simulated as a function of the steady-state light intensity $G_0$. In (b), a photo-induced voltage transient, under open-circuit conditions corresponding to 1 sun steady-state light intensity, is simulated. The inset depicts the voltage transient on a log-lin scale, with the related exponential fit indicated by the dashed line.

## 3. Comparison with Drift-Diffusion Simulations

To gain further insights into the physical meaning of the above theoretical analysis, we turn to transient device simulations based on a 1D drift-diffusion model.[37-39] The device model numerically solves Eq. (8) and (9), in conjunction with the Poisson equation and the time-dependent carrier continuity equations, assuming $J_c(x,t) = J_p(x,t) + J_n(x,t)$ to be given by the drift-diffusion relations.[40,41] In our simulations, an optically thin device with uniform photo-generation profiles (thus neglecting interference effects) is considered. The active layer is assumed undoped and have a thickness of $d = 100$ nm; moreover, balanced charge carrier mobilities of $10^{-4}$ cm$^2$/Vs are assumed. The recombination within the bulk is taken to be of a pure second-order type ($\mathcal{R} = \mathcal{R}_B = \beta np$), with a second-order recombination coefficient given by $\beta = 1.2 \times 10^{-11}$ cm$^3$/s. The electrical donor-acceptor bandgap of the active layer is set to 1.2 eV, whereas perfectly selective Ohmic contacts are assumed ($J_n(0) = J_p(d) = 0$), thus neglecting surface recombination effects[42]. The details of the simulations are given in the Supporting Information.

Figure 3a depicts the simulated current-voltage characteristics at 1 sun incident light intensity for the solar cell device considered, with pure second-order (bimolecular) recombination in the bulk being the dominant recombination mechanism. Ideally, we expect that $n_{\text{CE}} = \Delta n_{\text{eff}} = \sqrt{G_0/\beta}$ and $\tau_{TPV} = \tau_B = 1/(2\sqrt{\beta G_0})$, assuming equal and homogenous carrier distributions ($n = p$) under steady-state open-circuit conditions. In terms of the steady-state open-circuit voltage dependence, these ideal conditions are manifested by slope parameters of $m = 2$ and $\nu = 2$ for the extracted CE carrier density and TPV lifetime, respectively. These ideal conditions will accordingly also render the steady state parameter of the ideality factor $n_{id}$ equal to unity [see inset in Figure 3a]. In the following, the CE and TPV measurements are simulated and compared to the idealized case.

### 3.1. TPV - Capacitive Effects vs. Bulk Recombination

Figure 3b depicts a simulated TPV transient under 1 sun steady-state light intensity. The lifetime $\tau_{\text{TPV}}$ is extracted from the exponential decay of the simulated TPV transients (see inset of Figure 3b). The steady-state (dc) open-circuit voltage $V_{oc}$ is varied by varying the steady-state carrier photo-generation rate $G_0$; the simulated $V_{oc}$ as a function of $G_0$ is shown in the inset of Figure 3a. For simplicity the impact of the shunt resistance is assumed negligible, $R_{sh} \to \infty$.

In Figure 4a, the TPV lifetime $\tau_{\text{TPV}}$ as a function of the steady-state open-circuit voltage $V_{oc}$ is simulated for a load resistance of 1 GΩ, assuming a device area of $S = 0.1$ cm$^2$ (so that $R_L = 10^4$ Ωm$^2$). In accordance with Eq. (18), we expect the TPV lifetime to be determined by the steady-state values of $\partial n_{\text{eff}}/\partial \bar{\mathcal{R}}$ and $\partial J_D/\partial V$ at $V = V_{oc}$. Indeed, upon comparing the TPV lifetime extracted from the transient simulations (symbols) with Eq. (18), obtained from steady-state simulations (solid line), an excellent agreement is obtained. Furthermore, it can be seen that only at high light intensities (i.e. high $V_{oc}$), when the carrier profiles inside the active layer are sufficiently uniform, is the TPV lifetime accurately given by the bulk lifetime Eq. (4), as depicted by the dashed red line with the characteristic slope parameter of $\nu = 2$.

At smaller light intensities (lower $V_{oc}$), the TPV lifetime is governed by capacitive effects associated with back-injected dark carriers. In this regime, the TPV lifetime is very well approximated by Eq. (20) (depicted by blue short-dashed line in Figure 4a), assuming $C$ to be given by the dark dc capacitance (Figure S1). We note that this capacitance exhibits a voltage dependence under forward bias, increasing from $C \approx 1.1 C_0$ at $V = 0$ to $C \approx 1.4 C_0$ at $V = 0.5$ V, as can be seen from Figure S1. The increase in $C$, relative to $C_0$, can be understood in terms of back-injected dark carriers at the contacts penetrating deeper into the active layer, effectively decreasing the active layer thickness (in a geometric capacitor approximation). This capacitance is sometimes also referred to as the chemical or diffusion capacitance.[43-45]

It should be stressed that TPV is ultimately limited at lower voltages by the RC-times of the load resistance and/or the shunt resistance of the cell. The smaller of these two will set the RC-time constant of the system and defines the upper limit for the lifetime that can be extracted by the measurement. This is demonstrated in Figure 4b in case of a 1 MΩ load resistance (assuming $R_L \ll R_{sh}$); as $\tau_{\text{TPV}}$ approaches the RC-limit set by the load resistance, a saturation of the extracted lifetime $\tau_{\text{TPV}} \to R_L C$ will occur at low voltages, in accordance with both Eq. (18) and Eq. (20).

Based on these findings, it is only the TPV lifetimes at the highest intensities which is governed by recombination between photo-generated carriers inside the active layer, whereas the TPV lifetime at lower intensities is dominated by capacitive effects associated with spatially non-uniform charge carrier distributions within the bulk. Therefore, depending on the value of the shunt/load resistor, the TPV lifetimes at the lowest intensities are ruled only by the associated RC-time constant. Subsequently, a necessary requirement in TPV for the determination of the recombination lifetime is that $R_0 C_0 \ll \tau_{TPV} \ll [(1/R_L) + (1/R_{sh})]^{-1} C$.

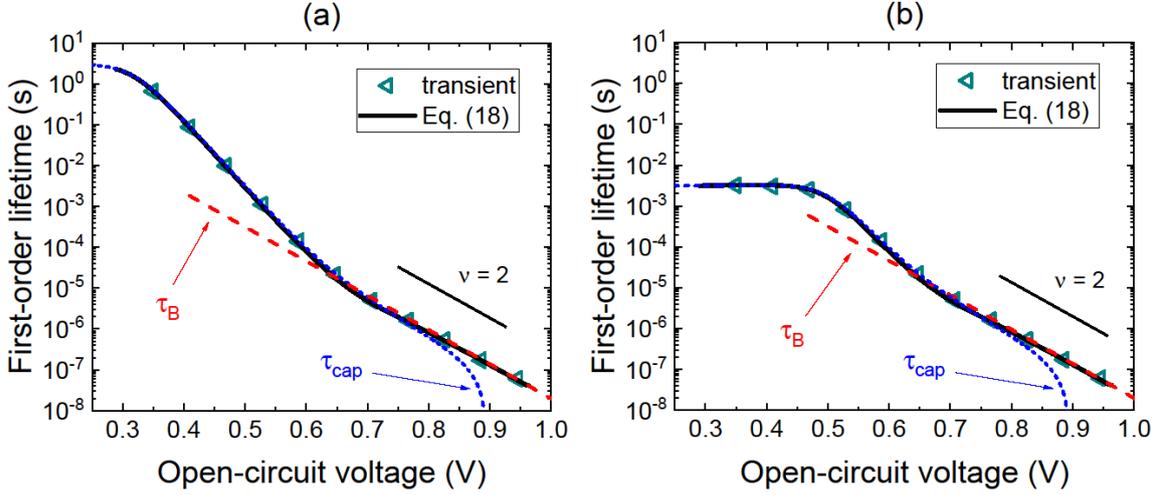

**Figure 4.** The TPV lifetime is shown as a function of the dc open-circuit voltage (steady-state light intensity) for two different load resistances: (a) 1 GΩ and (b) 1 MΩ. The lifetimes, as extracted from the simulated TPV transients (see inset of Figure 3b), are indicated by symbols, whereas Eq. (18) (derived from steady-state simulations) is depicted by the black solid line. The bulk lifetime ($\tau_B$) [Eq. (4)] and the capacitive RC ($\tau_{cap}$) [Eq. (20)] limits are indicated by the red dashed line and dotted blue line, respectively. A device area of $S = 0.1$ cm² is assumed.

## 3.2. The Validity of Complete Extraction and Uniform Carrier Distributions in CE

We next turn to charge extraction. The CE carrier density, shown in Figure 5, is given by $n_{CE} = Q_{CE}/qd$, with $Q_{CE}$ obtained by integrating the simulated extraction current transients and correcting for the geometric capacitance (see Eq. (22)). In Equation (25), it was assumed that (i) the recombination during the extraction process is negligibly small, and (ii) the carrier distributions are homogenous. To check the validity of these two assumptions, we have simulated CE current transients for the device considered in Figure 4a. The corresponding extracted carrier density $n_{CE}$ is shown in Figure 5a. This is to be compared to the excess carrier density $n_{oc} = \sqrt{G_0/\beta}$, as expected from Eq. (25), indicated by the red line (with a slope parameter of $m = 2$). It can be seen, for the case $\mu = 10^{-4}$ cm²/Vs, that Eq. (25) overestimates the actual $n_{CE}$ at higher $V_{oc}$, but underestimates it at low $V_{oc}$.

Figure 6 depicts the corresponding excess carrier densities $\Delta p(x)$ and $\Delta n(x)$, defined as the difference between the local carrier density at dc open-circuit conditions under illumination ($t = 0$) and in the dark ($t = \infty$), at a high and a low light intensity. **At high intensities**, shown in Figure 6a, the excess carrier profiles are indeed closely uniform inside the active layer, with the carrier density given by $n_{\text{oc}} = \sqrt{G_0/\beta}$. This suggests that assumption (ii) may be considered valid at high intensities. To check the validity of assumption (i), we compare the transient simulations (symbols) with Eq. (24) (solid lines), corresponding to the theoretical $n_{CE}$ expected in the case of complete charge extraction (as calculated from steady-state simulations) in Figure 5a. It can be seen that excellent agreement is obtained at small and moderate light intensities. Only at high intensities does a deviation occur, suggesting incomplete charge extraction at these intensities.

To investigate the reason for the incomplete charge extraction we compare the bulk lifetime Eq. (4) with the charge-carrier transit time, as shown in the inset of Figure 5a. The transit time, given by $t_{\text{tr}} \approx d^2/\mu V_{\text{oc}}$, corresponds to the time it takes a carrier to traverse the inter-electrode distance. It can be seen that for the simulated second order dynamics, the pseudo-first order bulk lifetime $\tau_B$ becomes smaller than $t_{\text{tr}}$ as $V_{\text{oc}}$ exceeds 0.8V resulting in a substantial recombination during the extraction process at larger intensities, explaining the incomplete charge extraction at these intensities. On the other hand, a much better agreement is obtained for the case with $\mu = 10^{-2}$ cm²/Vs, corresponding to a hundred times shorter transit time $t_{\text{tr}}$ (the other parameters are kept the same as before), in Figure 5a.

We can thus conclude that the underestimation of the carrier density seen at high intensities is related to the fact that higher-order (>1) recombination during the charge extraction process, starts to compete with extraction at higher intensities (see Eq. (4)). Therefore, a necessary condition to avoid recombination during the extraction pulse is that $t_{\text{tr}} \ll \tau_B$. This limitation can be potentially overcome by applying a strong reverse bias extraction voltage pulse, as shown by Kniepert et al,[46] although this will also give rise to a larger amount of displacement charges and thereby a larger noise on the extracted charge evaluation.

**At low light intensities** (small $V_{\text{oc}}$), the bimolecular recombination during the extraction process is negligible. However, as seen from Figure 6b, the assumption of uniform carrier distributions is no longer valid. Instead, the excess carrier distributions are highly non-uniform, varying exponentially with distance $x$ within the active layer, with the electron and hole profiles being mirror-symmetric to each other. This type of behaviour is generally expected for (dark) carriers, originating from the contacts.[47-51]

Accordingly, at low light intensity, the excess carrier densities of charges injected into the *undoped* active layer can be approximated by

$$\Delta n(x) \approx n_c \exp\left(-\frac{q[V_{bi}-V_{oc}]}{kT}\left[1-\frac{x}{d}\right]\right) \quad (26)$$

$$\Delta p(x) \approx p_a \exp\left(-\frac{q[V_{bi}-V_{oc}]}{kT}\frac{x}{d}\right) \quad (27)$$

for $V_{oc} < V_{bi}$, where $V_{bi}$ is an effective built-in potential that accounts for the energy-level bending in the vicinity of the injecting contacts,[52] and $p_a$ and $n_c$ are the associated effective hole and electron densities in the dark at the anode ($x = 0$) and cathode ($x = d$) regions, respectively. Then, by substituting Eq. (26) and (27) into Eq. (24) and integrating, we find

$$n_{CE} \approx \left(\frac{kT}{q[V_{bi}-V_{oc}]}\right)^2 [p_a + n_c] \quad (28)$$

when $(V_{bi} - V_{oc}) \gg kT/q$. In Figure 5b, we have compared Eq. (28) with the simulated CE data and a good agreement can found when assuming $p_a = n_c$ and $V_{bi}$ in Eq. (28) to be fixed and independent of the light intensity. In general, however, these quantities depend on the applied voltage (the injection level) as well; subsequently, a weak intensity dependence (via $V_{oc}$) of $p_a$, $n_c$ and $V_{bi}$ is to be expected. We note that these findings are consistent with previous results reported by Deledalle et al.[29]

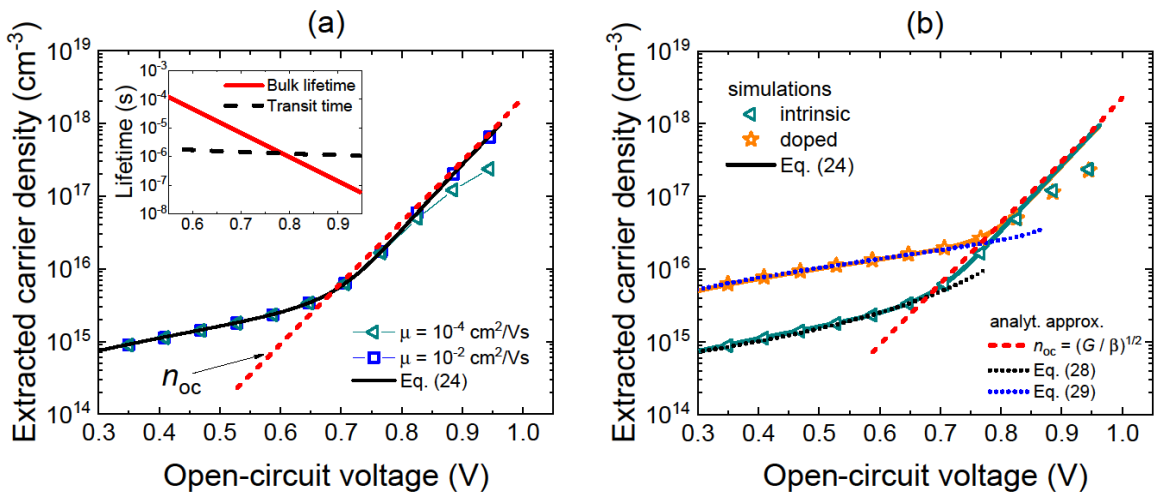

**Figure 5.** In (a), the extracted CE density $n_{CE} = Q_{CE}/qd$ (symbols), obtained from the simulated CE current transients, is shown as a function of open-circuit voltage for the organic solar cell for two different mobilities. The theoretical values in case of complete charge extraction Eq. (24), calculated from steady-state simulations, is indicated by solid lines. The expected analytical approximation Eq. (25) (i.e. $n_{CE} = n_{oc}$) is indicated by red dashed line. The inset shows the carrier transit time for $\mu = 10^{-4}$ cm²/Vs, compared to $\tau_B$ (Eq. (4)), at different $V_{oc}$. In (b), the transient simulations for the case with an undoped (intrinsic) and $p$-doped active layer are showed for the lower mobility case $\mu = 10^{-4}$ cm²/Vs. The solid lines depict the exact (numerical) Eq. (24) result, as calculated from steady-state simulations, for the two cases. The corresponding analytical approximations of Eq. (24) for the case with an undoped and a $p$-doped active layer at low intensities, given by Eq. (28) and (29), respectively, are indicated by dotted lines.

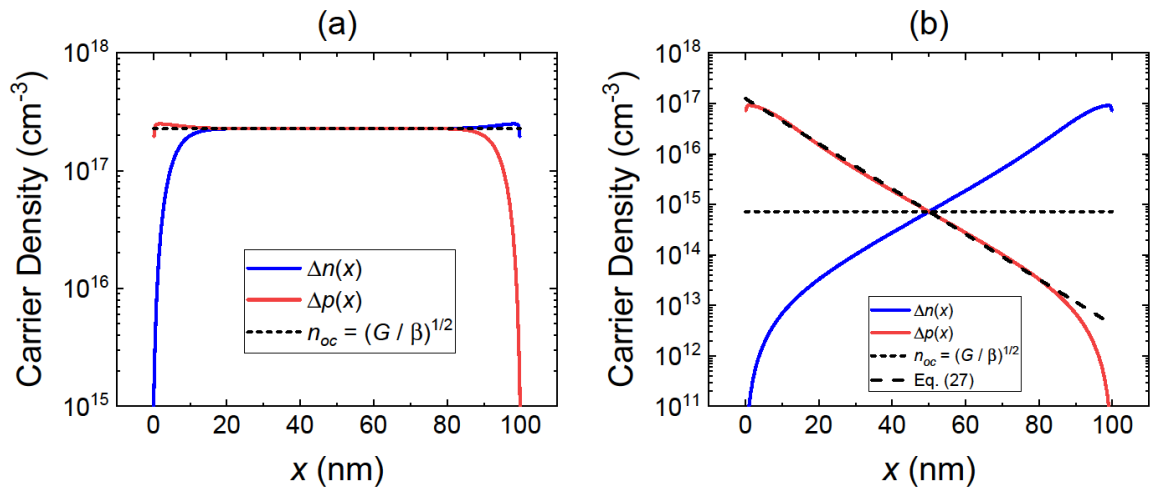

**Figure 6.** The simulated excess electron and hole densities, $\Delta n(x)$ and $\Delta p(x)$, respectively, as a function of distance $x$ inside the active layer. The excess carrier densities are given by the difference between the steady-state carrier density under open-circuit conditions under illumination ($t \leq 0$) and in the dark ($t \to \infty$); see Figure 2c. In (a), the situation at a high open-circuit voltage (high light-intensity regime, 100 suns) is simulated. The standard analytical approximation Eq. (25), assuming a uniform carrier profile given by $n_{oc} = \sqrt{G/\beta}$, is indicated by the dotted line. In (b), the situation at a lower open-circuit voltage (low light-intensity regime, 1/1000 suns) is simulated. For comparison, the exponential approximation Eq. (27) for $\Delta p(x)$ has been included (assuming $V_{bi} = 0.854$V and $p_a = 1.25 \times 10^{17}$ cm⁻³), as shown by the dashed line.

Our findings suggest that at low $V_{oc}$, the excess carrier density profiles are dominated by excess carriers, back-injected from the contacts, with $n_{CE}$ also being influenced by the associated capacitive effects ($\Delta \sigma'_{el} \neq 0$). To further validate that it indeed is the injected and/or dark background carriers that dominate the response in the capacitive CE regime, we have also included the case with a *p*-doped active layer in Figure 5b. In this case, the dark background carriers are dominated by doping-induced holes, with a carrier profile given by $p(x) \approx N_p$ for $x < d - w$, and $p(x) \approx 0$ else; where the depletion layer thickness is given by $w(V) \approx \sqrt{2\varepsilon\varepsilon_0[V_{bi} - V]/qN_p}$ and $N_p$ is the doping concentration.[40,44,53] Subsequently, Eq. (24) can be approximated as

$$n_{CE} \approx N_p \times \left[ \frac{w(0)}{d} - \frac{w(V_{oc})}{d} - \frac{1}{2}\left( \left[\frac{w(0)}{d}\right]^2 - \left[\frac{w(V_{oc})}{d}\right]^2 \right) \right] \quad (29)$$

which depends on $V_{oc}$ via $w$. Comparing Eq. (29) with the simulated CE data for the device with the doped active layer results in excellent agreement in the capacitive CE regime. These results also agree well with experimental data from Kiermasch et al.[24]

### 3.3. Discussion

Based on the recombination lifetime obtained from TPV, and the corresponding steady-state carrier density extracted with CE, the recombination rate and the lifetime as a function of carrier density is usually mapped.[13] The subsequent recombination rate, as derived from TPV and CE, can be expressed as

$$\mathcal{R}_{TPV/CE} = \frac{1}{1+\lambda} \times \frac{n_{CE}}{\tau_{TPV}} \quad (30)$$

where $\lambda = m/\nu$ and $1 + \lambda$ is the recombination order.

In line with the above findings, however, due to the inaccuracy of $n_{CE}$ at low light intensities, the TPV/CE recombination rate plotted against this extracted charge density is heavily plagued by capacitive effects. As discussed earlier this is caused by spatially separated carrier distributions and/or shunts. Only at high enough carrier densities, when $\tau_{TPV} \approx \tau_B$ and $n_{CE} \approx n_{oc}$, can the recombination rate constant be calculated from the recombination/charge plot. In this high carrier density regime, provided that the recombination during the extraction process in CE can be minimized, the slope of the recombination rate on a log-log plot, directly gives the associated recombination order.

We should note that here we considered balanced electron and hole mobilities and a relatively thin junction. Imbalanced mobilities or thicker junctions (which make the transit times of electrons and holes imbalanced) result in more problems in the validity of assumption (i), as one needs to consider the lifetime of the slower carriers and their transit time. On the other hand, in thicker devices, the photo-induced excess charge carrier profiles are generally more uniform throughout the active layer.[29] In addition, since capacitive effects can be reduced by increasing the active layer thickness, thicker devices are in general better suited for TPV but also for ensuring the validity of assumption (ii) in CE. This is consistent with the conclusions of previous work by Kirchartz and Deledalle.[19,29,54]

## 4. Experimental Demonstration on P3HT:PCBM Solar Cells

To demonstrate the relevance of the above theoretical findings, we turn to experimental results obtained from organic bulk heterojunction solar cells, based on P3HT:PCBM. In Figure 7a and Figure 7b the experimentally extracted TPV lifetime and CE charge carrier density, respectively, are shown as a function of the steady-state open-circuit voltage $V_{oc}$ of the P3HT:PCBM solar cell device in Reference 24. In Figure 7a, we have also included $\tau_{cap}$, as estimated from the experimental $J$-$V$ curve and the low-frequency capacitance of the device in the dark. Since this capacitance was found to show a relatively weak voltage dependence in the voltage range of interest, we assume a fixed value of $C = 0.55$ mF/m$^2$. Further experimental details of fabrication and measurements are outlined in Reference 24.

At small $V_{oc}$, good agreement is obtained between the estimated $\tau_{cap}$ and the experimental $\tau_{TPV}$ in this case. Conversely, at high enough $V_{oc}$, where we expect the recombination between bulk carriers to instead dominate the TPV response, a slope parameter of $\nu = 2$ is obtained. A similar situation is true for the CE data in Figure 7b, where $m = 2$ is obtained at high enough $V_{oc}$. The corresponding TPV/CE recombination rate, using the obtained slope $\lambda = 1$, is shown in Figure 7c as a function of $n_{CE}$. Accordingly, a slope of two is obtained in the relationship between total recombination rate and carrier density, i.e. $\mathcal{R}_{TPV/CE} \propto n_{CE}^2$, consistent with pure second-order recombination dominating at these light intensities. Furthermore, under these conditions (slope = 2), the measured data allows one to extract the second-order recombination coefficient via $\beta = (2n_{CE}\tau_{TPV})^{-1}$, in accordance with Eq. (4); from the $\beta$ vs $n_{CE}$ plot, depicted in Figure 7d, we find $\beta \approx 6 \times 10^{-12}$ cm$^3$/s.

The experimental results in this work suggests that the dominating recombination mechanism in P3HT:PCBM is second-order near 1 sun incident light intensities with a constant second-order recombination coefficient, being in line with previous findings by various other methods.[46,49,55-58] Furthermore, this also corroborates recent work by Tvingstedt and Deibel, who concluded that trap-assisted recombination via exponentially distributed tail states is not the dominating recombination mechanism in this system.[18]

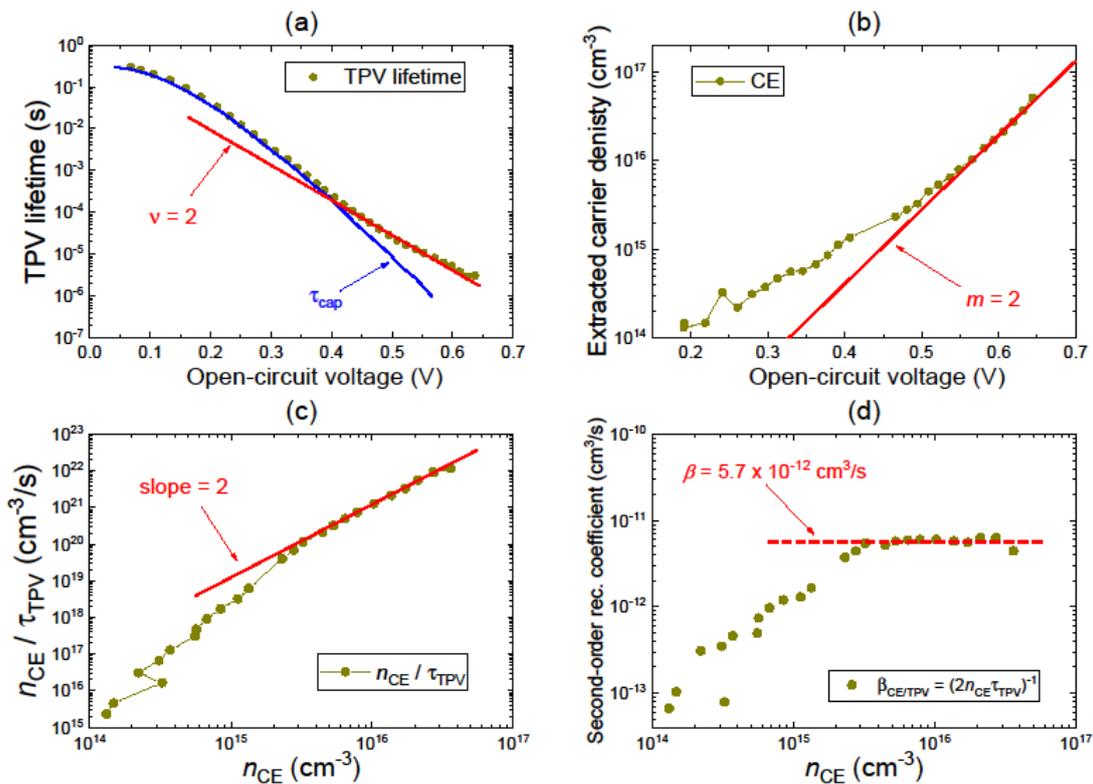

**Figure 7.** Experimental results on P3HT:PCBM bulk heterojunction solar cells. In (a), the experimental TPV lifetime, as indicated by the symbols, is shown. The corresponding $\tau_{cap}$, as estimated using the experimental $J$-$V$ curve, is shown by the dashed blue lines. (b) The experimental CE density $n_{CE} = Q_{CE}/qd$, where the extracted charge $Q_{CE}$ has been corrected for the capacitance. (c) The experimental CE-TPV recombination rate $n_{CE}/\tau_{TPV}$ is shown as a function of $n_{CE}$. It can be seen that at high intensities a slope of 2 is obtained, consistent with second-order recombination. (d) The corresponding second-order recombination coefficient $\beta_{CE/TPV} = 1/(2n_{CE}\tau_{TPV})$ is shown as function of $n_{CE}$.

## 5. Conclusions

In conclusion, based on the analytical derivations and numerical device simulations on realistic organic solar cells, the relation between the bulk recombination of photo-induced charge carriers and the lifetime, as extracted from the TPV transients, has been clarified. At higher $V_{oc}$ the lifetime is given by the sought-after bulk lifetime governed by bulk recombination. At lower light intensities and open-circuit voltages, the TPV lifetimes are instead limited by a composite RC-time constant which is dominated by the smallest of the load resistance of the measurement circuit, the shunt resistance (associated with parasitic leakage currents), and the internal (differential) resistance of the diode itself. For CE, the determination of the photo-generated carrier density $n_{oc}$ under open-circuit conditions is more challenging (and requires uniform carrier distributions) for higher order recombination. At higher $V_{oc}$, the CE measurement can be susceptible to recombination during the charge extraction process, which might lead to the extracted carrier density underestimating the actual steady-state density $n_{oc}$ prior to the pulse. A necessary requirement for the correct determination of $n_{oc}$ in this regime is that $t_{tr} \ll \tau_B$. At lower $V_{oc}$, however, the extracted CE density is dominated by capacitive effects caused by the extraction of spatially separated (back-injected) charge carriers. After accounting for the dark carrier profile (and their displacement current effects), analytical approximations for the capacitive CE regime could be obtained. Finally, the theoretical behaviour is reproduced experimentally on organic solar cells based on P3HT:PCBM.

**Associated Content**

Supporting Information: Analytical derivations, details regarding the device model, and additional simulations

**Acknowledgements**

The work was supported by the Sêr Cymru Program through the European Regional Development Fund, Welsh European Funding Office and Swansea University strategic initiative in Sustainable Advanced Materials. A.A. is a Sêr Cymru II Rising Star Fellow and P.M. a Sêr Cymru II National Research Chair. K.T. thanks the German Research Foundation (DFG) for funding through project 382633022 (RECOLPER).

TOC (embedded pdf)

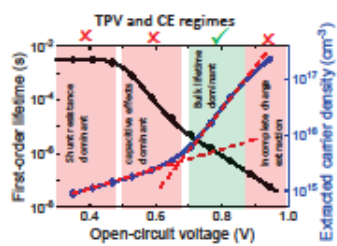